# *Ab initio* anharmonic thermodynamic properties of cubic CaSiO$_3$ perovskite

Zhen Zhang[1], Renata M. Wentzcovitch[1,2,3,*]

[1]*Department of Applied Physics and Applied Mathematics, Columbia University, New York, NY 10027, USA.*

[2]*Department of Earth and Environmental Sciences, Columbia University, New York, NY 10027, USA.*

[3]*Lamont–Doherty Earth Observatory, Columbia University, Palisades, NY 10964, USA.*

[*]To whom correspondence should be addressed.
rmw2150@columbia.edu

**Abstract**

We present an *ab initio* study of the thermodynamic properties of cubic $CaSiO_3$ perovskite (CaPv) over the pressure and temperature range of the Earth's lower mantle. We compute the anharmonic phonon dispersions throughout the Brillouin zone by utilizing the phonon quasiparticle approach, which characterizes the intrinsic temperature dependence of phonon frequencies and, in principle, captures full anharmonicity. Such temperature-dependent phonon dispersions are used to calculate *ab initio* free energy in the thermodynamic limit ($N \to \infty$) within the framework of the phonon gas model. Accurate free energy calculations enable us to investigate cubic CaPv's thermodynamic properties and thermal equation of state, where anharmonic effects are demonstrated. The present methodology provides an important theoretical approach for exploring phase boundaries, thermodynamic, and thermoelastic properties of strongly anharmonic materials at high pressures and temperatures.

## I. INTRODUCTION

CaSiO$_3$ perovskite (CaPv) is believed to be the third most abundant mineral in the Earth's lower mantle (LM), which constitutes 7 vol% of a pyrolitic LM [1,2]. As opposed to MgSiO$_3$ perovskite (MgPv) and MgO periclase (Pc), the first and second most abundant phases of the LM, of which thermodynamic [3,4] and thermoelastic [5,6] properties have been systematically investigated at high pressures (*P*) and high temperatures (*T*), CaPv's thermal properties have not been well characterized [7–10], mainly because of its strong anharmonicity. At low temperatures, e.g., *T* < 500 K [11,12], CaPv adopts a variety of tetragonal or orthorhombic phases [13]. At high temperatures, CaPv is dynamically stabilized by anharmonic fluctuations, and a cubic structure develops [11,12,14,15]. Although the exact *P-T* conditions under which the phase transition to cubic CaPv happens are still under debate [14,16], it is widely believed that under the LM conditions, 23 < *P* < 135 GPa and 2000 < *T* < 4000 K [5,17], the cubic phase with $Pm\bar{3}m$ space group is adopted [11,12].

Measurements of cubic CaPv's thermodynamic and thermoelastic properties are challenging because experiments are required to be performed at high *P-T* and cubic CaPv is unquenchable to ambient conditions [11]. Noguchi *et al.* [15] and Sun *et al.* [18] measured *P-V-T* data of cubic CaPv up to 150 GPa and 2600 K [18], but the compression curves under high *P-T* conditions have relatively large uncertainties. Recently, Gréaux *et al.* [19] and Thomson *et al.* [12] measured compressional wave ($v_p$) and shear wave ($v_s$) velocities of cubic CaPv up to 23 GPa and 1700 K [19], whereas their reported thermoelastic parameters under LM conditions are based on extrapolations. *Ab initio* studies of cubic CaPv's thermodynamic and thermoelastic properties also encounter difficulties. The quasiharmonic approximation (QHA), which has been successfully applied to MgPv and Pc [3–6], is invalid for cubic CaPv because of the presence of unstable phonon normal modes with imaginary frequencies at all pressures using harmonic phonon calculations [7,11]. Kawai and Tsuchiya [20,21] conducted *ab initio* molecular dynamics (MD) simulations to study the thermodynamic and thermoelastic properties of cubic CaPv. However, whether the thermodynamic properties [20] are fully converged is questionable due to the finite-size effect inherent in the MD approach [11]. A more in-depth and systematic investigation of cubic CaPv's thermal properties is required for a complete understanding of the dynamic state of the deep Earth.

In this study, we report the anharmonic thermodynamic properties of cubic CaPv at LM conditions by using our well-established phonon quasiparticle approach [22]. The phonon quasiparticle approach is a hybrid approach combining *ab initio* lattice dynamics and MD simulations, which fully accounts for anharmonic effects. In other words, it treats phonon anharmonicity to all orders in principle. It has been successfully applied to strongly [11,23] and weakly [22] anharmonic systems, metallic [23] and non-metallic [11,22] anharmonic systems, and used to compute anharmonic phonon dispersions [11,22–24], anharmonic free energies [11,22], a pre-melting phase transition [23], and lattice thermal conductivities [25,26]. Here the phonon quasiparticles of cubic CaPv are first extracted from mode-projected velocity autocorrelation function (VAF) obtained by *ab initio* MD simulations. Next, the phonon quasiparticle frequencies, known as renormalized frequencies, are Fourier interpolated over the Brillouin zone (BZ). Then the thermodynamic properties in the thermodynamic limit ($N \to \infty$) are obtained within the framework of the phonon gas model (PGM) [27,28].

The PGM always serves as a paradigm in calculating the thermodynamic properties of crystalline materials, which uses the phonon spectrum to compute vibrational entropy, free energy, and, thus, thermodynamic quantities. For weakly anharmonic systems, a commonly used simplification of the PGM is the QHA, which neglects the intrinsic temperature dependence of phonon frequencies and treats phonon frequencies as explicitly volume-dependent only. The QHA works well for materials of this class because it accounts for the extrinsic temperature dependence of the phonon frequencies caused by volume change. The QHA fails in strongly anharmonic systems, whereas the PGM in general still applies as long as the phonon quasiparticles are well-defined, i.e., with well-defined frequencies and lifetimes [29]. The *ab initio* MD-based direct free energy method, e.g., thermodynamic integration (TI) [30], is another widely used method in dealing with strong anharmonicity. However, to approach the thermodynamic limit, conducting TI using *ab initio* MD with a sufficiently large supercell is beyond the current computational capability. The advantage of the present PGM approach is that it uses the phonon quasiparticle spectrum obtained on a sufficiently large **q**-mesh to compute well converged thermodynamic quantities [11,22]. Furthermore, here we correct the total energy error originating in the density functional theory (DFT) calculations [31,32] by making reference to previously reported experimental *P-V-T* data [15,19] of cubic CaPv.

## II. METHOD

In the present approach, a phonon quasiparticle of normal mode (**q**, *s*) is numerically defined by the VAF [11,22],

$$\langle V_{\mathbf{q}s}(0) \cdot V_{\mathbf{q}s}(t) \rangle = \lim_{\tau \to \infty} \frac{1}{\tau} \int_0^\tau V_{\mathbf{q}s}^*(t') V_{\mathbf{q}s}(t' + t) dt', \quad (1)$$

where $V_{\mathbf{q}s}(t) = \sum_{i=1}^N V(t) \cdot e^{i\mathbf{q} \cdot \mathbf{r}_i} \cdot \hat{\mathbf{e}}_{\mathbf{q}s}$ is the (**q**, *s*)-mode-projected velocity. **q** is the phonon wave vector, and *s* labels the 3*n* phonon branches of an *n*-atom primitive cell. $V(t) = V\left(\sqrt{M_1}\mathbf{v}_1(t), \ldots, \sqrt{M_N}\mathbf{v}_N(t)\right)$ is the mass-weighted velocity with 3*N* components, where $\mathbf{v}_i(t)(i=1,\ldots,N)$ is atomic velocity produced by *ab initio* MD simulations of an *N*-atom supercell, and $M_i$ is the atomic mass of the $i^{th}$ atom in the supercell. $\hat{\mathbf{e}}_{\mathbf{q}s}$ is the harmonic phonon polarization vector of mode (**q**, *s*), which is calculated by the density functional perturbation theory (DFPT) [33]. For a well-defined phonon quasiparticle, its power spectrum,

$$G_{\mathbf{q}s}(\omega) = \left| \int_0^\infty \langle V_{\mathbf{q}s}(0) \cdot V_{\mathbf{q}s}(t) \rangle e^{i\omega t} dt \right|^2 \quad (2)$$

should have a Lorentzian-type line shape with a peak at $\widetilde{\omega}_{\mathbf{q}s}$ and a phonon linewidth of $1/(2\tau_{\mathbf{q}s})$ [22,29], $\widetilde{\omega}_{\mathbf{q}s}$ being the (**q**, *s*)-mode renormalized frequency and $\tau_{\mathbf{q}s}$ being the lifetime. Phonon lifetimes can be used to investigate the lattice thermal conductivity [25,26]. Here we rely on the renormalized frequencies to compute anharmonic thermodynamic properties.

As reported by the previous studies, the effective harmonic dynamical matrix can be constructed as [11,22,23],

$$\widetilde{D}(\mathbf{q}) = [\hat{\mathbf{e}}_{\mathbf{q}}] \Omega_{\mathbf{q}} [\hat{\mathbf{e}}_{\mathbf{q}}]^\dagger, \quad (3)$$

where the diagonal matrix $\Omega_{\mathbf{q}} = \text{diag}[\widetilde{\omega}_{\mathbf{q}1}^2, \widetilde{\omega}_{\mathbf{q}2}^2, \ldots, \widetilde{\omega}_{\mathbf{q}3N}^2]$ contains $\widetilde{\omega}_{\mathbf{q}s}^2$ in the diagonal, and $[\hat{\mathbf{e}}_{\mathbf{q}}] = [\hat{\mathbf{e}}_{\mathbf{q}1}, \hat{\mathbf{e}}_{\mathbf{q}2}, \ldots, \hat{\mathbf{e}}_{\mathbf{q}3N}]$ is the matrix of harmonic eigenvectors. The effective harmonic force constant matrix, $\widetilde{\Phi}(\mathbf{r})$, can be obtained from the Fourier transform of $\widetilde{D}(\mathbf{q})$, where the anharmonic interaction is effectively captured. Therefore, $\widetilde{\omega}_{\mathbf{q}'s}$ at any wave vector $\mathbf{q}'$ in the BZ can be obtained by diagonalizing,

$$\widetilde{D}(\mathbf{q}') = \sum_{\mathbf{r}} \widetilde{\Phi}(\mathbf{r}) \cdot e^{-i\mathbf{q}' \cdot \mathbf{r}}, \quad (4)$$

from which the anharmonic phonon dispersion and vibrational density of states (VDoS) at finite temperatures are computed.

Here, *ab initio* MD simulations were carried out in the *NVT* ensemble using the DFT-based Vienna *ab initio* simulation package (VASP) [34] employing the local density approximation (LDA) and the projected-augmented wave method (PAW) [35]. Cubic CaPv was simulated with a $2 \times 2 \times 2$ (40 atoms) supercell with adopting a shifted $2 \times 2 \times 2$ **k**-mesh and a kinetic energy cutoff of 550 eV. Our previous study has shown that $2 \times 2 \times 2$ supercell of CaPv is sufficient to converge anharmonic interaction and anharmonic phonon dispersion [11], since anharmonic parts of interatomic forces have shorter ranges than the harmonic ones. MD simulations were conducted on a series of volumes ($V$), 44.39, 40.26, 36.77, 34.34, and 32.49 Å$^3$/primitive cell, corresponding to densities ($\rho$), 4.35, 4.79, 5.25, 5.62 and 5.94 g/cm$^3$, respectively. The temperature ranging from 1500 to 4000 K was controlled by Nosé thermostat [36]. Each simulation ran for over 60 ps with a time step of 1 fs. Harmonic phonon normal modes were calculated using DFPT [33] implemented in the VASP package. Throughout the $V$, $T$ range investigated, phonon quasiparticles were well-defined, and the cubic phase of CaPv was confirmed.

**III. RESULTS AND DISCUSSION**

The renormalized phonon frequencies, $\widetilde{\omega}_{\mathbf{q}s}$, are first extracted from phonon quasiparticles sampled by the MD simulations. In order to converge thermodynamic properties, it is desirable to obtain $\widetilde{\omega}_{\mathbf{q}s}$ on a much denser **q**-mesh to approximate the thermodynamic limit. Eq. (4) enable us to obtain $\widetilde{\omega}_{\mathbf{q}s}$ at any **q**-point throughout the BZ, hence the anharmonic phonon dispersion and VDoS [11,22–24]. The obtained temperature-dependent phonon dispersions of cubic CaPv at $\rho$ = 5.25 g/cm$^3$ are showcased in Fig. 1(a), where the intrinsic temperature-dependence of $\widetilde{\omega}_{\mathbf{q}s}$ in the BZ is clearly exhibited. The corresponding temperature-dependent VDoS obtained on a $20 \times 20 \times 20$ **q**-mesh approximating the thermodynamic limit are shown in Fig. 1(b). Except for the acoustic branches centered at wave vector $\mathbf{R}(\frac{1}{2},\frac{1}{2},\frac{1}{2})$, which only accounts for a small portion of the phonon dispersion, frequencies of most phonon modes are weakly temperature-dependent. This is counterintuitive since CaPv is strongly anharmonic. $\widetilde{\omega}_{\mathbf{q}s}$ of most phonon branches show nonmonotonic temperature-dependence, while only optical modes with $\widetilde{\omega}_{\mathbf{q}s}$ above ~800 cm$^{-1}$ display discernible frequency shift down with increasing temperature. Nevertheless, such temperature-dependence is comparable to that of MgPv, which is weakly anharmonic [22,25], at the same *P-T* conditions [26]. The anharmonic phonon dispersions are further used to compute vibrational entropy ($S$) and free energy ($F$) within the framework of the PGM [27,28].

When using the temperature-dependent phonon dispersions to compute thermodynamic properties, the QHA free energy formula is no longer valid. Nevertheless, the entropy formula [11,22,23,29],

$$S(T) = k_B \sum_{\mathbf{q}s}[(n_{\mathbf{q}s} + 1)\ln(n_{\mathbf{q}s} + 1) - n_{\mathbf{q}s}\ln n_{\mathbf{q}s}], \qquad (5)$$

where $n_{\mathbf{q}s} = [\exp(\hbar\widetilde{\omega}_{\mathbf{q}s}(T)/k_B T) - 1]^{-1}$, is still applicable. $\widetilde{\omega}_{\mathbf{q}s}(T)$ at arbitrary temperatures were obtained by fitting a second-order polynomial in $T$ to $\widetilde{\omega}_{\mathbf{q}s}$ [22,23] calculated at several temperatures and constant volume. The obtained $S(T)$ at different densities are shown in Fig. 2(a). The Helmholtz free energy can be obtained by [23],

$$F(V,T) = E(V,T_0) - T_0 S(V,T_0) - \int_{T_0}^{T} S(T')dT', \qquad (6)$$

where the reference temperature $T_0$ = 1500 K, and $E(V,T_0)$ is the time-averaged internal energy obtained from the MD simulation at $T_0$. The obtained $F(V)$ at different temperatures are shown as dashed curves in Fig. 2(b). The present PGM approach relying on renormalized phonon frequencies to compute vibrational entropy, free energy and thermodynamic quantities overcomes the deficiencies of QHA in dealing with strongly anharmonic materials in two aspects. First, the intrinsic anharmonic effects arise from phonon-phonon interaction, and the phonon frequency is explicitly temperature-dependent instead of implicitly dependent on volume only. Second, the crystal structure is stabilized by anharmonic fluctuations only at high temperatures while being dynamically unstable at low temperatures. Our previous studies have shown that the present PGM approach gives consistent anharmonic entropy with that provided by TI using the same supercell [11,22]. Besides, by Fourier interpolating the renormalized phonon frequencies over the BZ, the PGM also overcomes the finite-size effect inherent in TI [11,22], the simulation cell size of which is limited by the computational capability of *ab initio* MD. The difference between well converged free energy in the thermodynamic limit and the one obtained from a finite size supercell is significant for determining phase boundaries [22,23].

Cubic CaPv's isothermal equations of state (EoS) are computed by fitting the $F(V,T)$ to a third-order finite strain expansion at each temperature. For practical applications of these results to Earth's interior modeling, errors in the total energy originating in the exchange-correlation functional used, the LDA, and possibly also in the PAWs adopted are undesirable. Here we introduce an additional correction to $F(V,T)$ to bring the calculated EoS into full agreement with experimentally measured high-temperature EoSs [12,15,18,19]. Anharmonicity, in principle, is adequately addressed by the quasiparticle approach. To obtain theoretical isothermal compression curves in good agreement with experiments, it is desirable to provide proper constraints at both low and high pressures. Here we adopted Gréaux *et al.*'s [19] and Noguchi *et al.*'s [15] experimental *P-V-T* data for cubic CaPv to impose such constraints at low and high pressures, respectively. As for other recently reported experimental results, Thomson *et al.* [12] conducted measurements up to ~16 GPa and therefore provides the same constraint of the compression curve by Gréaux *et al.* [19] at low pressures. Sun *et al.*'s [18] and Noguchi *et al.*'s [15] measurements have relatively significant uncertainties in pressure. Noguchi *et al.* [15] performed both laser heating and external heating diamond-anvil-cell (DAC) experiments, and their data are consistent with Gréaux *et al.*'s [19] data obtained in multi-anvil. The experimental data by Sun *et al.* [18] were obtained using a laser-heated DAC, while multi-anvil and resistance-heated DAC should have better temperature control. Therefore, Sun *et al.*'s [18] measurements were not used in the energy correction procedure. The reference temperature, $T_{ref}$, chosen to make the correction was 1600 K since experimental *P-V* data is available near this temperature. Noguchi *et al.*'s measurements were conducted at ~1600 K [15] and Gréaux *et al.*'s measurements at ~1500 and ~1700 K [19]. The calculated compression curve was corrected by adopting the generalized Kunc-Syassen scheme (KSr) [31,32],

$$\Delta V(P) = \frac{V_0^{exp}}{V_0^{DFT}} V\left(\frac{K_0^{DFT}}{K_0^{exp}} P, K'_{exp}\right) - V_{DFT}(P), \quad (7)$$

where $V_0^{exp}$, $K_0^{exp}$ and $K'_{exp}$ are parameters obtained from measurements at $T_{ref}$, while $V_0^{DFT}$, $K_0^{DFT}$ and $K'_{DFT}$ are parameters obtained from $F(V,T)$ at the same $T_{ref}$. $\Delta V(P)$ can then be easily inverted to give $\Delta P(V)$. In this way, the correction to $F(V,T)$ at $T_{ref}$ can be obtained as $\Delta F(V) = \int \Delta P(V) dV$. Once the $F(V)$ was corrected at $T_{ref}$, the same correction $\Delta F(V)$ was then applied to other temperatures, i.e., 1500 K < *T* < 4000 K. The corrected $F(V)$ at different temperatures are shown as solid curves in Fig. 2(b). By fitting the corrected $F(V,T)$ to a third-order finite strain expansion at each temperature, the resulting pressure-volume EoS isotherms are shown as solid curves in Fig. 3, along with uncorrected ones shown as dashed curves. The corrected EoS are in good agreement with measured data within experimental uncertainties. At $T_0$ = 1500 K, the EoS parameters obtained are $V_0$ = 46.39 Å³/primitive cell, $K_{T0}$ = 264 GPa, and $K'_{T0}$ = 3.1, where $V_0$ is the equilibrium volume, $K_{T0}$ is the isothermal bulk modulus at $V_0$, and $K'_{T0}$ is the pressure derivative of the isothermal bulk modulus at $V_0$, respectively.

With $V(P,T)$ and $P(V,T)$ obtained, cubic CaPv's thermal expansivity ($\alpha$) and isothermal bulk modulus ($K_T$) are readily calculated as,

$$\alpha = \frac{1}{V}\left(\frac{\partial V}{\partial T}\right)_P, \tag{8}$$

and,

$$K_T = -V\left(\frac{\partial P}{\partial V}\right)_T. \tag{9}$$

The obtained $\alpha(T)$ and $K_T(T)$ at a series of LM pressures are displayed in Figs. 4(a) and 4(b), respectively. Compared with $\alpha$ of MgPv [3] and Pc [4] obtained using the QHA, the inclusion of intrinsic anharmonic effects for cubic CaPv gives a slow and approximately linear temperature-dependence of $\alpha$ at low pressures [4]. With increasing pressure, $\alpha$ decreases, and the effects of temperature become less and less pronounced, resulting essentially in temperature-independent $\alpha$ at constant pressure. The $\alpha$ by Kawai and Tsuchiya 2014 [20] is also shown for comparison. Compared with our results before EoS correction, their $\alpha$ has a rather rapid temperature-dependence and is larger than ours at low pressures and high temperatures. The overestimation of $\alpha$, especially at low pressures and high temperatures, is an indication of the inadequacy of the QHA [4,37]. Hence it is likely also a hint of not fully accounting for anharmonic effects in the work by Kawai and Tsuchiya 2014 [20]. Apart from the anharmonic effects, the EoS correction also contributes to the discrepancy in $\alpha$ between the present study and Kawai and Tsuchiya 2014 [20].

The thermodynamic Grüneisen parameter ($\gamma$) is a very important quantity often used to quantify the relationship between thermal and elastic properties. It is defined as

$$\gamma = \frac{V\alpha K_T}{C_V}, \tag{10}$$

where $C_V$ is the isochoric heat capacity. It is also a useful indicator of the importance of anharmonicity, increasing with the latter [4]. The calculated $\gamma(T)$ at a series of pressures are shown in Fig. 5. Similar to $\alpha$, cubic CaPv's $\gamma$ is nearly independent of temperature at all LM pressures. In some cases, $\gamma$ of CaPv slowly decreases with temperature, as opposed to the monotonically increasing behavior reported by Kawai and Tsuchiya 2014 [20], and those of MgPv [3] and Pc [4] obtained by QHA. Unlike $\alpha$, the discrepancy in $\gamma$ between our study and Kawai and Tsuchiya 2014 [20] originates mainly in whether anharmonicity being fully accounted for. The EoS correction plays a minor role. The volume dependence of $\gamma$ is often expressed by a parameter, $q =$ $(\partial \ln \gamma / \partial \ln V)_T$. At $P$ = 20 GPa and $T$ = 1500 K, we find $q$ = 0.80. At 20 GPa and with increasing temperature, $q$ decreases to 0.38 at 4000 K. At 1500 K, with increasing pressure, $q$ decreases to -0.16 at 140 GPa.

$C_V$ of cubic CaPv is calculated from temperature-dependent anharmonic phonon dispersions by,

$$C_V = T\left(\frac{\partial S}{\partial T}\right)_V, \qquad (11)$$

within the PGM. Fig. 6(a) compares the $C_V$ calculated this way accounting for full anharmonicity, with that derived from temperature-independent anharmonic phonon dispersion obtained only at the reference temperature $T_0$ = 1500 K. With increasing temperature, the latter $C_V$ converges to the Dulong-Petit classical limit, $3nk_B$, which is also the high-temperature limit within the Debye model [38] for harmonic crystal with temperature-independent phonon frequencies. While the former $C_V$ can surpass such limit because of anharmonicity [39,40]. The anharmonic contribution to CaPv's $C_V$ at constant volume increases nearly linearly with temperature [39]. Hence, $C_V$ is another important and straightforward indicator of anharmonic effects inherent in the phonon frequencies. The isobaric heat capacity ($C_P$) is given by

$$C_P = C_V(1 + \gamma \alpha T), \qquad (12)$$

the temperature and pressure dependence of which are summarized in Fig. 6(b). $C_P$ is valuable for experimentally determining CaPv's lattice thermal conductivity, $\kappa = D\rho C_P$, where $D$ is the measured thermal diffusivity.

The adiabatic bulk modulus ($K_S$) is related to the isothermal one ($K_T$) by,

$$K_S = K_T(1 + \gamma \alpha T). \qquad (13)$$

The obtained $K_S(T)$ at different pressures are displayed in Fig. 7(a), compared with a previous study by Kawai and Tsuchiya 2014 [20]. At all pressures, $K_S$ is nearly temperature-independent and slightly decreases with temperature. The $K_S(P)$ along several isotherms are shown in Fig. 7(b), compared with previous studies by Kawai and Tsuchiya 2015 [21], and Thomson *et al.* [12]. Thomson *et al.* conducted measurements for cubic CaPv's $v_p$ and $v_s$ in a narrow pressure range, i.e., up to ~16 GPa, and extrapolated its thermoelastic properties to LM conditions using literature *P-V-T* data [12]. The different experimental literature data used in the present study [15,19] and Thomson *et al.* [12,15,18,41,42] results in a discrepancy in the pressure dependence of $K_S$. Besides, the present study makes DFT energy correction with reference to the literature data at $T_{ref}$ = 1600 K, while anharmonicity is adequately addressed by the phonon quasiparticle approach at higher temperatures. Note that Kawai and Tsuchiya 2014 [20] obtained $K_S$ in the thermodynamic way via Eq. (13), while Kawai and Tsuchiya 2015 [21] obtained $K_S$ by calculating the thermoelastic parameters, which slightly differs from the former. The discrepancy in $K_S$ between our study and Kawai and Tsuchiya 2014 [20] results from the lack of EoS correction using experimental P-V-T data by them and the accumulated differences in $\alpha$ and $\gamma$ accounting for anharmonicity. As for Kawai and Tsuchiya 2015, our calculated $K_S$ before EoS correction agrees with their results relatively well, meaning the anharmonic effects on $K_S$ are properly addressed by Kawai and Tsuchiya 2015 [21]. The discrepancy between our corrected results and theirs originates in our EoS correction process.

The Mie-Grüneisen EoS [43,44] is a commonly used relation to determine the high-temperature pressure in shock-compressed solids. It can be described as,

$$P(V,T) = P_0(V) + P_{th}(V,T), \quad (14)$$

where $P_0$ is the pressure at the reference temperature, $T_0$, and $P_{th}$ is the thermal pressure. In principle, $P_0$ is well described by the third-order Birch-Murnaghan EoS,

$$P_0(V) = \tfrac{3}{2} K_{T0} \left[ \left(\tfrac{V}{V_0}\right)^{-7/3} - \left(\tfrac{V}{V_0}\right)^{-5/3} \right] \left\{ 1 + \tfrac{3}{4}(K'_{T0} - 4)\left[\left(\tfrac{V}{V_0}\right)^{-2/3} - 1\right] \right\}. \quad (15)$$

$P_{th}$ is expressed by the difference of thermal energy, $E_{th}$, between $T$ and $T_0$,

$$P_{th}(V,T) = \tfrac{\gamma_{mg}(V)}{V}[E_{th}(V,T) - E_{th}(V,T_0)], \quad (16)$$

where $\gamma_{mg}$ is the Mie-Grüneisen EoS Grüneisen parameter. $E_{th}$ is related to the Debye temperature, $\theta$,

$$E_{th}(V,T) = 9nRT\left(\tfrac{\theta(V)}{T}\right)^{-3} \int_0^{\theta(V)/T} \tfrac{x^3}{e^x - 1} dx, \quad (17)$$

where $n$ is the number of atoms per formula unit, and $R$ is the gas constant. $\gamma_{mg}(V)$ is expressed as,

$$\gamma_{mg}(V) = \gamma_{mg0}\left(\tfrac{V}{V_0}\right)^{q_{mg}}, \quad (18)$$

where $q_{mg}$ is a volume-independent parameter. $\theta(V)$ is expressed as,

$$\theta(V) = \theta_0 \exp\left(-\tfrac{\gamma_{mg} - \gamma_{mg0}}{q_{mg}}\right), \quad (19)$$

where $\gamma_{mg0}$ and $\theta_0$ are the Mie-Grüneisen EoS Grüneisen parameter and Debye temperature at $(V_0, T_0)$, respectively.

Here we chose $T_0$ = 1500 K and adopted the obtained isothermal EoS parameters, $V_0$, $K_{T0}$ and $K'_{T0}$. Then we fit the calculated *P-V-T* data at higher temperatures to the Mie-Grüneisen EoS relation to obtain the remaining EoS parameters, $\theta_0$, $\gamma_{mg0}$ and $q_{mg}$. We found $\gamma_{mg0}$ and $q_{mg}$ to be insensitive to the variation of $\theta_0$, which is consistent with previous reports [15,18]. Also, it is a common practice to fix $\theta_0$ [18,20,41,42], and fit for $\gamma_{mg0}$ and $q_{mg}$. Therefore, we first evaluated $\theta_0$ from the Debye model [38,45],

$$\theta = \frac{h}{k_B}\left(\frac{3nN_A\rho}{4\pi M}\right)^{1/3} v_m, \quad (20)$$

where $h$, $N_A$ and $M$ are Plank constant, Avogadro number, and molecular mass per formula unit. $v_m$ is the average wave velocity integrated over several crystal directions [45],

$$v_m = \left[\frac{1}{3}\left(\frac{1}{v_p^3}+\frac{2}{v_s^3}\right)\right]^{-1/3}. \quad (21)$$

Here we adopted $v_p$ = 9.28 km/s and $v_s$ = 5.17 km/s of cubic CaPv at 0 GPa and 1500 K from Gréaux *et al.* [19]. The resulting $\theta_0$ is 815 K. Finally, by fixing $\theta_0$ and allowing $\gamma_{mg0}$ and $q_{mg}$ to vary, we obtained the fitting parameters $\gamma_{mg0}$ = 1.49, and $q_{mg}$ = 0.68. The obtained Mie-Grüneisen EoS parameters perfectly describe the calculated *P-V-T* data shown in Fig. 3, which are summarized in Table I and compared with several previous studies [12,15,18,20,41,42]. We did not express the Mie-Grüneisen EoS at low reference temperatures, e.g., 300 K, for two reasons. First, cubic CaPv is unquenchable to ambient conditions and unstable at low temperatures. Second, anharmonicity addressed by phonon quasiparticles cannot be extrapolated to low temperatures at which quasiparticles are not well-defined and the stable structure is different.

An interesting fact to note is that the thermodynamic Grüneisen parameter, $\gamma$, defined by Eq. (10) and displayed in Fig. 5, differs from the Mie-Grüneisen EoS Grüneisen parameter, $\gamma_{mg}$. For example, at 1500 K and 30 GPa, our calculated $\gamma$ = 1.48, while $\gamma_{mg}$ = 1.40. At 1500 K and 140 GPa, $\gamma$ = 1.24, while $\gamma_{mg}$ = 1.20. The two quantities coincide with each other [43,44] when satisfying three criteria. First, the system is within the framework of QHA, so that $\gamma$ can be approximated by $\bar{\gamma}$ [43,44],

$$\bar{\gamma} = \frac{\sum_i \gamma_i C_{V_i}}{C_V}, \quad (22)$$

where $C_{V_i}$ is the mode isochoric heat capacity, and $\gamma_i = -(\partial \ln \omega_i / \partial \ln V)$ is the mode Grüneisen parameter. Meanwhile, $\gamma_{mg}$ is associated with $\gamma_i$ [43,44],

$$\gamma_{mg} = \frac{\sum_i \gamma_i E_{th_i}}{E_{th}}, \tag{23}$$

where $E_{th_i}$ is the mode thermal energy. Second, by assuming all $\gamma_i$ are equal to each other [44], $\gamma_i$ is factored out of Eq. (22) and (23), in which way $\gamma = \bar{\gamma} = \gamma_{mg}$. However, realistically, $\gamma_i$ are not equal. Therefore third, only at sufficiently high temperature, e.g., all $C_{V_i}$ are equal and all $E_{th_i}$ are equal, $\gamma_{mg}$ is an approximation to $\bar{\gamma}$ [44]. Here for cubic CaPv, none of the criteria is satisfied, resulting in a difference between $\gamma$ and $\gamma_{mg}$.

## IV. CONCLUSIONS

In summary, we have computed the temperature-dependent anharmonic phonon dispersions of cubic CaPv throughout the Earth's lower mantle conditions using the phonon quasiparticle approach. Anharmonic phonon dispersions with stable phonons enabled us to evaluate the *ab initio* free energy, $F(V,T)$, in the thermodynamic limit ($N \to \infty$) [11,22] within the phonon gas model [27,28]. DFT energy errors were corrected by carefully combing [31,32] calculated $F(V,T)$ and pressure, $P(V,T)$, with experimental *P-V-T* data [15,19]. The corrected $F(V,T)$ was used to investigate the cubic CaPv's thermal equation of state (EoS) and several thermodynamic quantities of interest. The calculated thermal expansivity and thermodynamic Grüneisen parameter show nearly temperature-independent behavior, which is a sign of the importance of anharmonic effects [4]. The intrinsic temperature dependence of phonon frequencies caused by phonon-phonon interaction leads to a discernibly larger isochoric heat capacity beyond the classical limit at high temperatures [39,40]. The calculated *P-V-T* data are also fit to the Mie-Grüneisen EoS. The obtained Mie-Grüneisen EoS Grüneisen parameter differs [43,44] from the thermodynamic Grüneisen parameter, which is, in part, also caused by anharmonicity. The present approach for accurate free energy calculations can be applied to investigate phase boundaries [23] and thermodynamic and thermoelastic properties of other strongly anharmonic systems at high pressures and temperatures.

## ACKNOWLEDGMENTS

This work was primarily funded primarily by the US Department of Energy Grant DE-SC0019759 and in part by the National Science Foundation (NSF) award EAR-1918126. This work used the Extreme Science and Engineering Discovery Environment (XSEDE), USA, which was supported by the NSF Grant ACI-1548562. Computations were performed on Stampede2, the flagship supercomputer at the Texas Advanced Computing Center (TACC), the University of Texas at Austin generously funded by the NSF through Grant ACI-1134872.

TABLE I. Mie-Grüneisen EoS parameters of this study compared with previous studies [12,15,18,20,41,42].

|  | This study | Wang et al. | Shim and Duffy | Noguchi et al. | Kawai and Tsuchiya | Sun et al. | Thomson et al. |
|---|---|---|---|---|---|---|---|
| $T_0$ (K) | 1500 | 300 | 300 | 700 | 1000 | 300 | 300 |
| $V_0$ (Å$^3$) | 46.39 | 45.58 | 45.58 | 46.5 | 46.17 | 45.4 | 45.57 |
| $K_{T0}$ (GPa) | 264 | 232 | 236 | 207 | 203.5 | 249 | 248 |
| $K'_{T0}$ | 3.1 | 4.8 | 3.9 | 4 | 4.76 | 4 | 3.6 |
| $\theta_0$ (K) | 815 | 1100 | 1000 | 1300 | 1100 | 1000 | 771 |
| $\gamma_{mg0}$ | 1.49 | 1.7 | 1.92 | 2.7 | 1.576 | 1.8 | 1.67 |
| $q_{mg}$ | 0.68 | 1.0 | 0.6 | 1.2 | 0.96 | 1.1 | 1.1 |

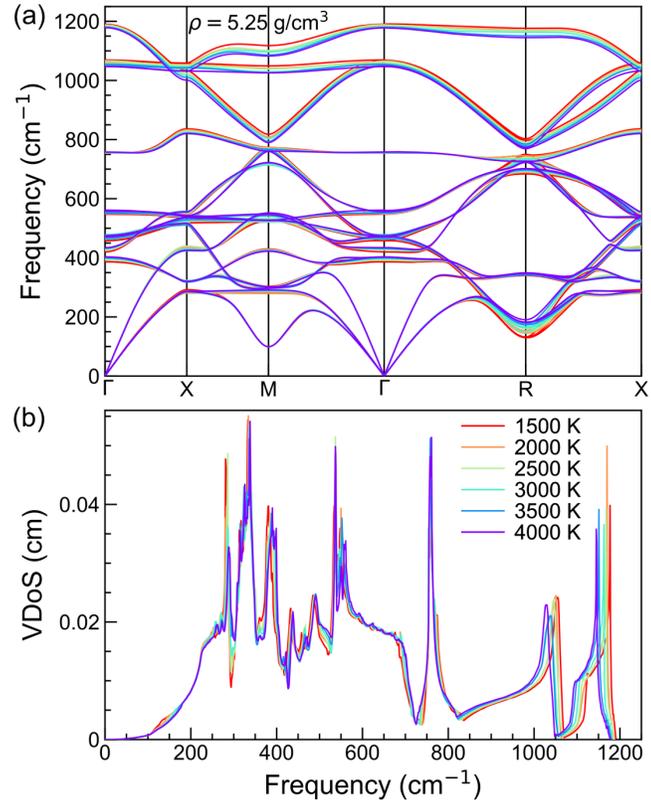

FIG. 1. Anharmonic (a) phonon dispersions and (b) vibrational density of states (VDoS) at a series of temperatures at constant density.

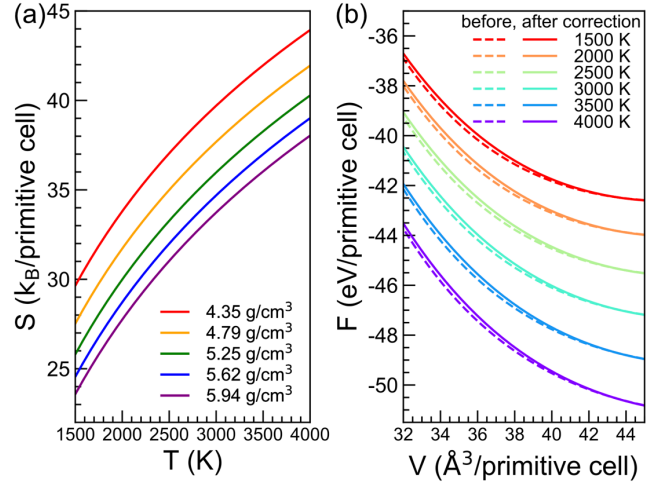

FIG. 2. (a) Vibrational entropy ($S$) vs. $T$ at a series of densities. (b) Helmholtz free energy ($F$) vs. $V$ at a series of temperatures before (dashed curves) and after (solid curves) DFT energy correction.

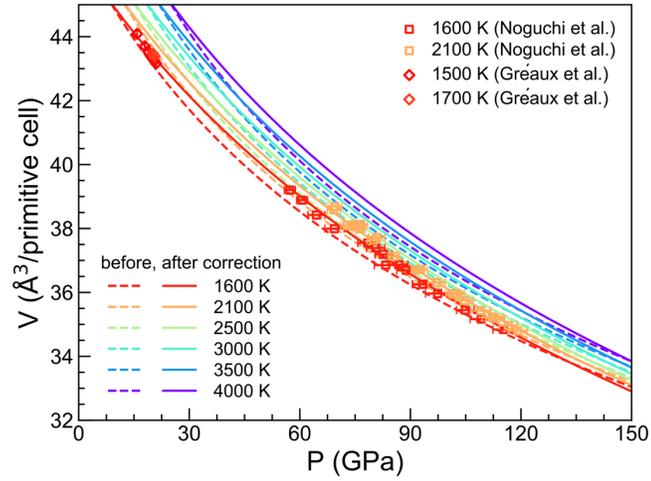

FIG. 3. Isothermal third-order finite strain EoS at a series of temperatures before (dashed curves) and after (solid curves) DFT energy correction, compared with experimental measurements [15,19]. Error bars show the experimental uncertainties.

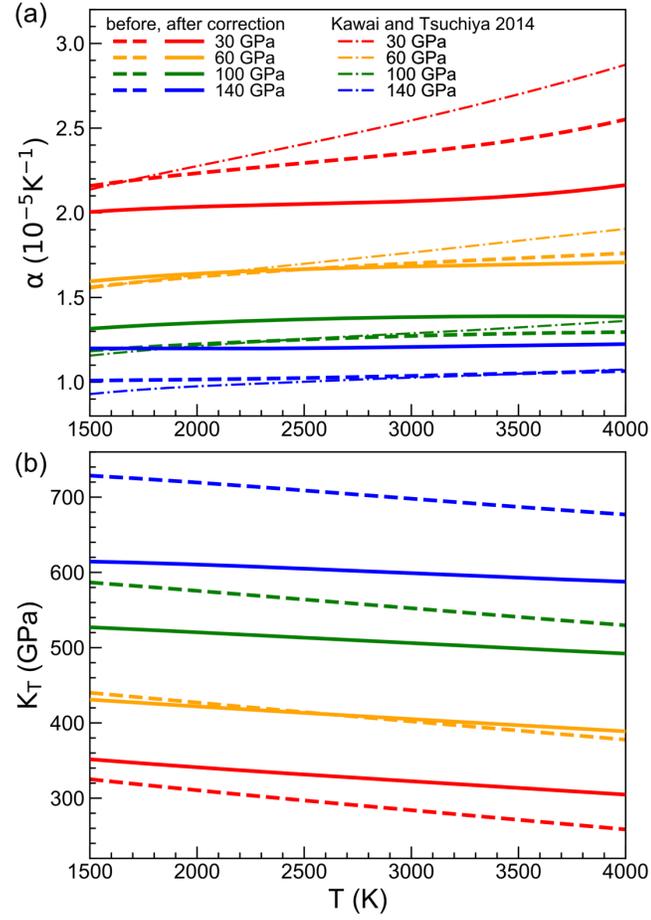

FIG. 4. (a) Thermal expansivity ($\alpha$) and (b) isothermal bulk modulus ($K_T$) vs. $T$ at a series of pressures before (dashed curves) and after (solid curves) DFT energy correction. Dash-dotted curves are results from a previous study [20].

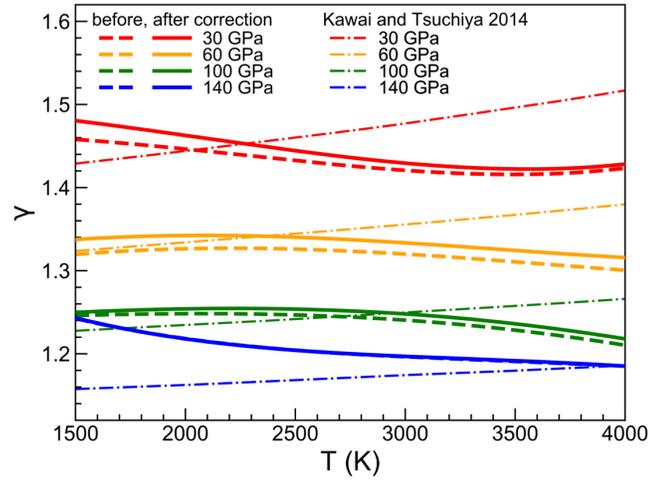

FIG. 5. Thermodynamic Grüneisen parameter ($\gamma$) vs. $T$ at a series of pressures before (dashed curves) and after (solid curves) DFT energy correction. Dash-dotted curves are results from a previous study [20].

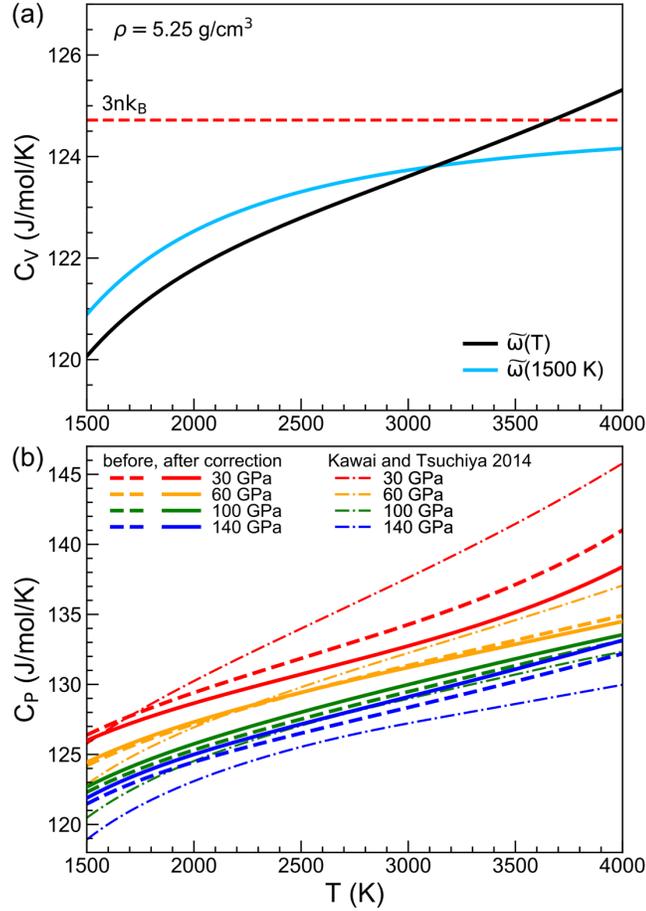

FIG. 6. (a) Isochoric heat capacity ($C_V$) vs. $T$ at constant density. Solid black curve was calculated from temperature-dependent anharmonic VDoS, and solid blue curve was calculated from temperature-independent anharmonic VDoS obtained only at the reference temperature $T_0 = 1500$ K. Red dashed line labels the classical limit, $3nk_B$. (b) Isobaric heat capacity ($C_P$) vs. $T$ at a series of pressures before (dashed curves) and after (solid curves) DFT energy correction. Dash-dotted curves are results from a previous study [20].

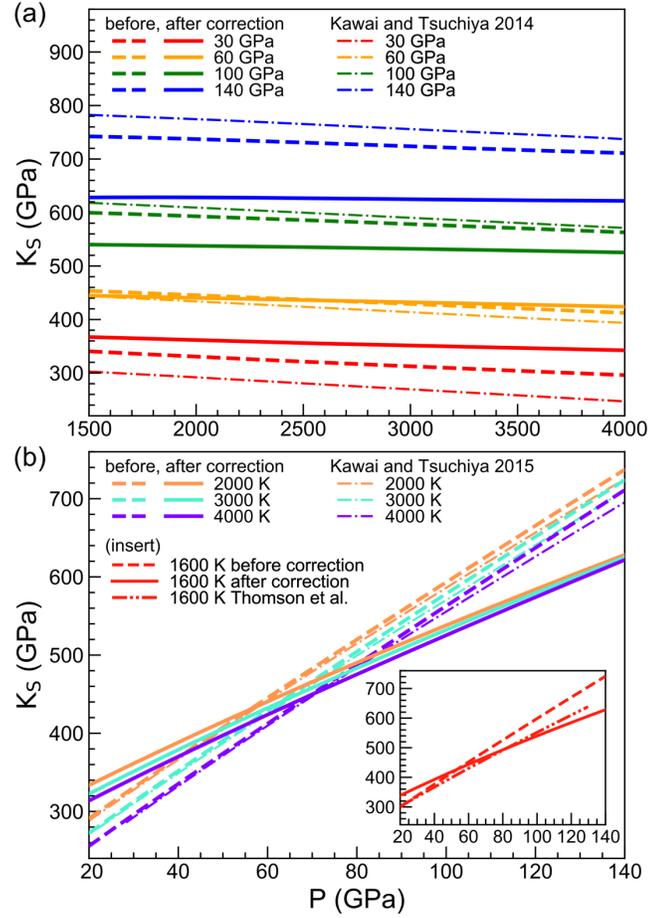

FIG. 7. (a) Adiabatic bulk modulus ($K_S$) vs. $T$ at a series of pressures before (dashed curves) and after (solid curves) DFT energy correction. (b) $K_S$ vs. $P$ along several isotherms. Dash-dotted and dash-dot-dotted curves are results from previous studies [12,20,21].